\documentclass[letterpaper]{article}
\usepackage{iccc}
\usepackage{graphicx}
\usepackage{natbib}
\usepackage{hyperref} 
\usepackage{dblfloatfix}
\usepackage{times}
\usepackage{helvet}
\usepackage{courier}
\title{Yourfeed: Towards open science and interoperable systems for social media}

\author{Ziv Epstein\textsuperscript{1*}, Hause Lin\textsuperscript{2,3*}\\
\textsuperscript{1}MIT Media Lab, \textsuperscript{2}MIT Sloan, \textsuperscript{3}University of Regina\\
\{zive, hause\}@mit.edu\\
\textsuperscript{*}These authors contributed equally
}
\begin{document} 
\maketitle
\begin{abstract}
\begin{quote}
Existing social media platforms (SMPs) make it incredibly difficult for researchers to conduct studies on social media, which in turn has created a knowledge gap between academia and industry about the effects of platform design on user behavior. To close the gap, we introduce Yourfeed, a research tool for conducting ecologically valid social media research. We introduce the platform architecture, as well key opportunities such as assessing the effects of exposure of content on downstream beliefs and attitudes, measuring attentional exposure via dwell time, and evaluating heterogeneous newsfeed algorithms. We discuss the underlying philosophy of interoperability for social media and future developments for the platform.
\end{quote}
\end{abstract}

\section{Introduction}
Today, more than half of the population of earth are on social media. As a tool for individuals to gain information about the world and connect with others, and as a platform for researchers to build a generalizable understanding about human nature \citep{lazer2020computational}, the possibilities for social media to provide meaningful information and facilitate collective action are nearly endless. Yet the corporatized modern platforms that exist today have in many ways fallen short of this lofty vision. 

For one, the profit incentive of social media platforms (SMPs) has prioritized designs and interfaces that put the priorities of the company above the user \citep{lorenz2020behavioural}. Newsfeed algorithms optimized for engagement (e.g. Facebook \citep{fb_wpo}) or dwell (e.g. TikTok \citep{tiktok_nyt, tiktok_wsj}) keep users on-platform and in turn may drive the spread of misinformation \citep{hao2021facebook}, while many users are unaware of the fact that algorithms are dictating what they see \citep{eslami2015always}. Dark patterns (UX/UI elements that subtly trick you \citep{gray2018dark}, for examples see \cite{harris}) abound on user interfaces. Platforms control user data and content policies, with little transparency or recourse for how these policies are created and enforced \citep{sadowski2021everyone}. 

In addition, the culture and structure of social media platforms have drastically limited researchers' ability to use them to generate public knowledge about their workings \cite{greene2022barriers, bail2022social}. Internally, SMPs have sophisticated experimental frameworks for understanding the effect of algorithms and interfaces on user behavior, which they use to further refine their platforms for profit. Yet these insights are held behind walled gardens of NDAs and industry secrets. Sometimes, the results of these studies are publicly shared, such as a 61-million person experiment to increase voter turnout \citep{bond201261}, or the infamous experiment of emotional contagion \citep{kramer2014experimental}. In both cases, there was major public backlash about the impact of these systems, and the questionable ethics of such experiments \citep{jouhki2016facebook, zittrain2014facebook, zuboff2015big}. 

For academics attempting to ask similar questions, they must rely on crude alternatives. One is partnerships with the companies, where the SMP grants the researchers access to user data to write public facing papers. However to ensure privacy, these datasets are often aggregated to a point that they are of limited use, and these partnerships might stall for political reasons \citep{ss1}. Furthermore, the SMP providing the data can set the agenda, and differentially favor academics whose research interests align with their strategy \citep{hegelich2020facebook}. Another approach is social media field experiments, where academics attempt to leverage platform APIs to conduct experiments with real users \citep{mosleh2022field}. These field experiments can however violate platforms Terms and Services, ``poison the well'' by undermining trust in digital ecosystems \citep{pennycook2021shifting}, and are often challenging to replicate as platforms change their policies to adapt. A final approach is controlled laboratory experiments with survey tools like Qualtrics. This paradigm allows researchers to run RCTs to understand the effect of design interventions on user behavior (see e.g. \cite{pennycook2021psychology}). However, the context of Qualtrics leaves a lot to be desired for drawing generalizable inferences about the impact of interfaces and algorithms on user behavior \citep{lin2021promises}. In particular, the standard Qualtrics approach presents social media posts one at a time, rather than in a newsfeed layout. This feed layout, however, is a critical design feature of modern social media platforms, and therefore Qualtrics omits key attentional features of online social media use and consumption. 
\begin{figure*}[hbp]
\includegraphics[width=0.99\textwidth]{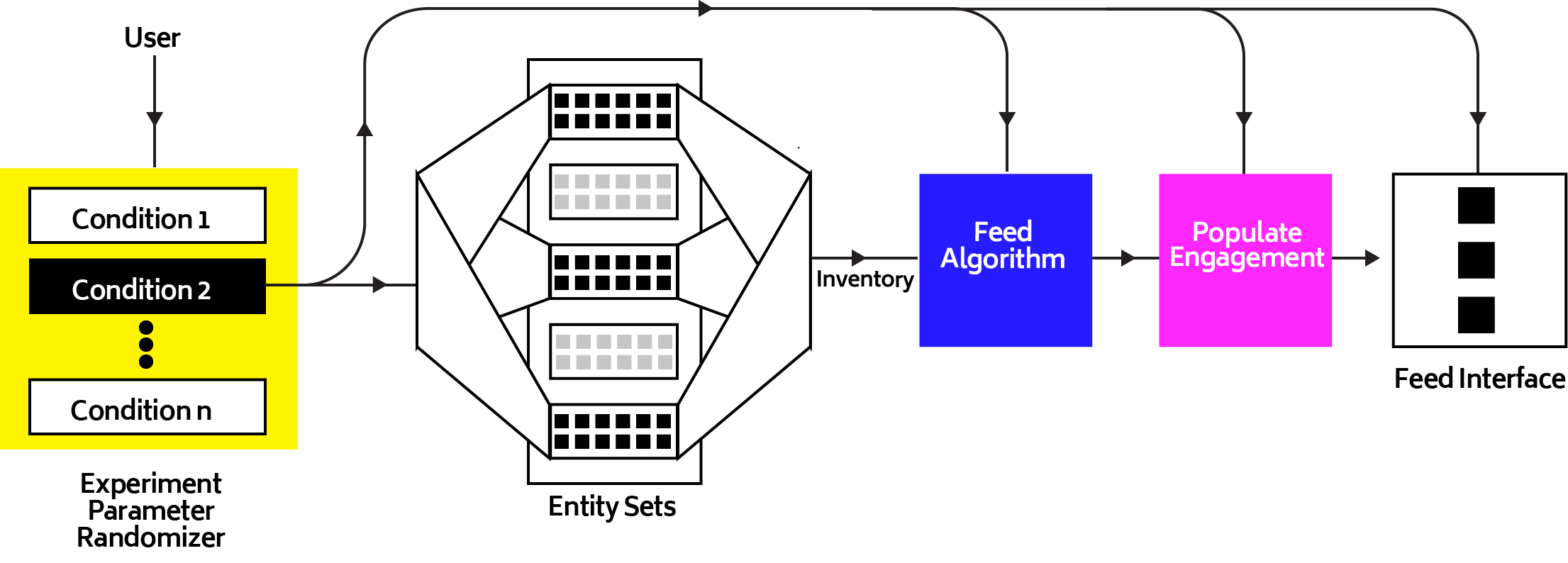}
\centering
\caption{Overview of the platform architecture}
\label{fig:architecture}
\end{figure*}

To close the knowledge gap between the platforms and the public, we introduce Yourfeed, a modular tool for ecologically valid social media research. Yourfeed mirrors the attentional properties of social media by providing a feed layout for content. In addition, it tracks and surfaces the dwell time of each post, a key metric from industry that is hard to measure on traditional survey platforms like Qualtrics. Yourfeed is designed to be modular and interoperable, allowing researchers to upload their own components, such as stimulus sets, ranking algorithms, or design interventions. In this paper, we introduce the Yourfeed system, with several key features and studies researchers could run. We conclude with the implications of Yourfeed on building public knowledge about platforms impact user behavior, and how this knowledge might help nudge platforms to open up their policies and surfaces for a more democratic approach.

\begin{figure*}[h!]
\includegraphics[width=0.99\textwidth]{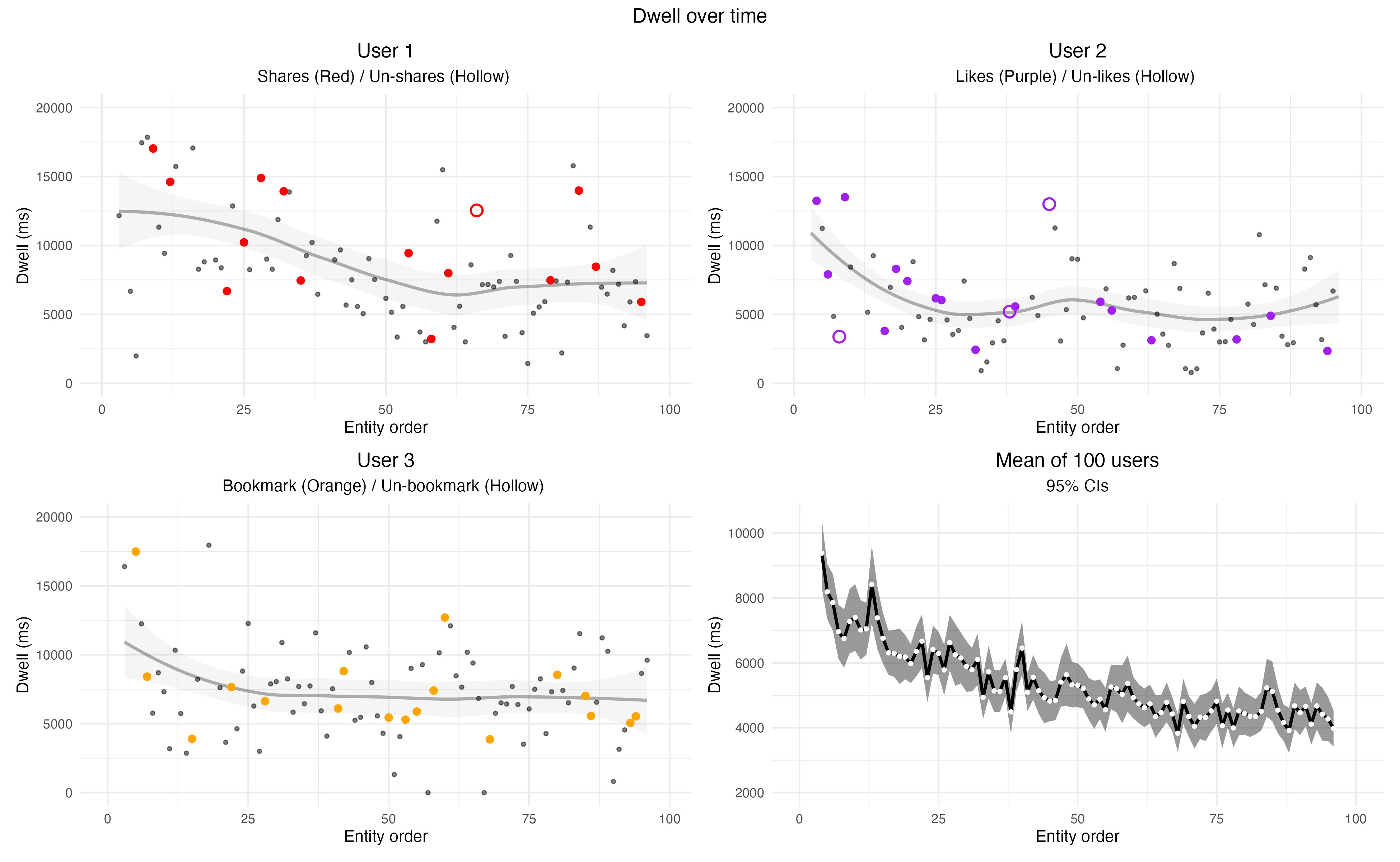}
\centering
\caption{Top left: Dwell and share data for a user as they scroll through Yourfeed. A red dot corresponds to posts the user decided to share, while a hollow red dot corresponds to posts the user shared and then unshared. Top right: Dwell and like data for a user as they scroll through Yourfeed. Bottom left: Dwell and bookmark data for a user as they scroll through Yourfeed. Bottom right: Average dwell time of 100 users over the course of a browsing session (Entity order = 0 is the top of the newsfeed, entity order=100 is the bottom).}
\label{fig:dwell}
\end{figure*}
\section{System Overview}
Yourfeed is an experimental tool that allows researchers to conduct randomized studies in a feed environment (see Figure~\ref{fig:architecture}). To do so, researchers specify a series of conditions (e.g. 1 to n) on the create-studies page and a series of attributes for each condition. Once a set of experimental parameters has been submitted, Yourfeed generates a unique URL for that specific experiment. When a user first visits Yourfeed at that URL, the Experiment Parameter Randomizer assigns that user to a condition (the yellow box in Figure~\ref{fig:architecture}), and then renders the feed interface using the attributes for that condition. In particular, there are several types of attributes used to generate the feed interface for a given user. The first are the entity sets, which are sets of newsfeed items to display in the feed. Experimenters can select these sets for each condition, and also specify how many items from a given entity set are displayed. For each user, the items from the specified entity sets constitute that user’s inventory. This inventory is then ordered by a feed algorithm, which can be a user-specified blackbox algorithm (the blue box in Figure~\ref{fig:architecture}). The default algorithm simply orders the feed items in a random order, but researchers can upload their own bespoke algorithms in a standardized format. The sorted inventory is then populated with engagement scores (e.g. the number of likes or posts, the pink box in Figure~\ref{fig:architecture}). These engagement scores could be omitted if specified by the experiment parameters, sampled randomly for each post for each user (e.g. as down in \cite{avram2020exposure}), or actually show the number of likes or shares from other participants in the study. Finally, the sorted inventory populated with engagement is then displayed in the feed interface, which is additionally styled according to experimental parameters (e.g. with advertisements, a modal intervention, or a Instagram/Facebook skin). Following the feed interaction, participants proceed to a survey environment where post-task demographics and beliefs/attitudes can be assessed. Researchers can also upload their own entity sets at the upload page, and download the data for their experiments at the download-data page. The platform is available online at \url{https://www.yourfeed.social}.

The affordances of Yourfeed allow researchers to ask research questions that would be difficult otherwise. Below, we enumerate several key features of Yourfeed and example research questions and studies that can be conducted with those features. 

\subsection{More ecologically valid studies on social media behavior and attitudes}
There are many studies researchers can conduct on survey platforms like Qualtrics that can be directly transported into Yourfeed with added ecological validity. One class of studies involves measuring the causal effect of exposure of content on downstream beliefs and attitudes. Since researchers can specify different entity sets for different conditions, they can randomly assign participants to exposure of different topics, then follow up with the same set of survey questions as dependent variables. For example, a researcher might randomly vary the amount of content related to climate change, and look at how that influences how important participants perceive climate change to be. A second class of studies involves measuring the impact of a design intervention on sharing behavior. For example, a researcher might introduce an interstitial accuracy prompt at a random point in the feed, and look how this interstitial impacts sharing discernment (both the magnitude and duration of the treatment effect).

An important implication of Yourfeed’s design is that share rates on the platform are lower. In a recent Yourfeed experiment with 876 participants recruited from Prolific, the overall average share rate was 7.7\%, overall average like rate was 12.1\%, compared to a share rate of 32.5\% and a like rate of 31.\% on Qualtrics (using data from \cite{epstein2021social}). This suggests that share rates from Qualtrics, while correlated with actual sharing \citep{mosleh2020self}, may be systematically larger than they would be on platform.

\subsection{Quantifying attention dynamics via dwell time}
A key limitation of survey experiments on engagement with social media is that they often show social media posts one by one, which forces attentional capture on each one. Such setups are contrived and differ dramatically from the design of feed-based platforms, where some posts grab attention while others are scrolled past. Building on the metrics and practices of the online advertising industry \citep{hwang2020subprime}, Yourfeed measures the amount of time a user spends looking at a particular post by recording the amount of time that post is viewable (e.g. dwell time). A researcher could treat this value itself as an outcome variable, and look at which user- and item-level covariates are predictive of different attention dynamics. 

To underscore the kinds of data Yourfeed collects and generates, we present pilot data from a study conducted on Yourfeed (run in June 2022 with N=876 participants recruited from Prolific). As shown in Figure~\ref{fig:dwell}, Yourfeed provides granular data about how long each participants dwelled on a given post (entity), and how they engaged with each. We also find that dwell time decreases over a browsing session, whereby participants spend the most time on the top item in their feed, and the least on the bottom.

\subsection{Serving and evaluating heterogeneous newsfeed algorithms}
Recently, Gizmodo made public a series of Facebook papers which elucidate the inner workings of how Facebook operates. Among these are experiments that compare algorithmic ranking to chronological newsfeeds \citep{gizmodo}. There has been growing concern about the effects of algorithmic ranking on user behavior \citep{eckles2022algorithmic, sassine2022influence}, and political information flow \citep{reuning2022facebook} but a lack of methods for studying these effects directly has resulted in leaked internal SMP documents being a key method for public knowledge on the topic. 

Just as entity sets or design interventions can be randomly assigned to participants, so can feed algorithms. This feature allows researchers to causally assess the effects of different algorithms on user behavior. For example, a growing body of work has recommended newsfeed algorithms that are more aligned with social priorities \citep{cen2021regulating, bhadani2022political}, and Yourfeed provides a platform for evaluating the effects of these algorithms on user behavior. 

In addition to randomly assigning participants to different algorithms, Yourfeed could randomly assign participants to different worlds, each of which evolve independently \citep{salganik2006experimental, epstein2021social}. This feature allows researchers to quantify the variation introduced by such algorithms via counterfactuals, and to measure ecosystem-level outcomes, like the diversity of content \citep{epstein2021social}.

As an example, a researcher could create a study with three conditions: a control condition where items are randomly ordered, a second condition with a newsfeed that optimizes and sorts by engagement (e.g. shares, like Facebook), and third condition with a newsfeed that optimizes by dwell time (like TikTok). For the two algorithmic conditions, participants could be assigned to one of 10 worlds which evolve independently. In particular, for a given user the order of the newsfeed is determined solely by the patterns of engagement (in the second condition) or dwell (in the third condition) by other participants assigned to that user’s world. The researcher could compute the diversity of content across each world, and therefore assess the causal effect of the ranking algorithms on diversity across worlds.

\section{Related Work}
Many recent papers have developed custom platforms to explore behavioral aspects of social media. \citet{avram2020exposure} developed Fakey, a simulated newsfeed game that over 8,500 users played. They randomized the engagement metrics associated with each post, and find that exposure to social engagement metrics increases vulnerability to misinformation. \citet{roozenbeek2019fake} developed Bad News, a browser based that over 15,000 users engaged with. They found that users’ ability to detect misinformation increases after gameplay, across demographic subgroups. \citet{epstein2021social} developed Meet the Ganimals, an online platform for generating and curating AI-generated hybrid animals \citep{epstein2020interpolating}, that over 10,000 players engaged with. OpenHuman \citep{greshake2019open} is a community-based platform that gives users higher levels of personal data access and control.

Other experimental platforms have explored participatory algorithm design and algorithmic auditing. \citet{bhargava2019gobo} developed Gobo, an experimental system for exploring user control of invisible algorithms in social media. Gobo allowed users to control the ranking of posts in their newsfeed with manual controls, such as the seriousness or rudeness of posts, or the gender of the poster. Turingbox \citep{epstein2018turingbox} was an experimental two-sided market where contributors could upload algorithms for evaluations, and social scientists could evaluate them, towards understanding the consequences of their behavior \citep{epstein2018turingbox}. 

Perhaps most closely related to Yourfeed are tools developed for researchers to conduct social media research. The Mock Social Media Website Tool \citep{msmwt} is an open-source software for studying social media. Participants in the MSMW can like posts, share posts on a timeline, add their own posts and reply to other posts. The (Mis)information Game \citep{misinfo_game} is an online testing platform that allows researchers to customize posts, source information and engagement information. Unlike a feed enviornment, (Mis)information Game presents posts one at  time for users to interact with and comment on. Community Connect \citep{mahajan2021community} is an open-source and customizable platform for experimentation, designed for controlled information flow across groups through bridge nodes.
\section{Discussion}
There are several components that constitute modern social media platforms. The first is the underlying social data on which the platform is based, such as the social graph and posts or comments. The second is the recommender algorithm which for a given user’s inventory serves a subset of that content to their feed. The third and final is a “skin” or interface by which this feed content is displayed and users interact with the medium. The current model of social media offers a monolithic experience with each of these components molded together, with no opportunities for users to select alternative components, like newsfeed algorithms or skins. This walled-garden approach means that SMPs can carefully design these components with their incentives in mind (e.g. maximizing profit via engagement optimization) and serve them to users who have no choice but use them if they want to access the platforms. Yourfeed offers an alternative, interoperable vision of social media. Rather than a cloister, there could exist a rich ecology of third-party newsfeed algorithms and skins, each designed with a different perspective and goal in mind. Users could access the same underlying data, but specify the particular components that satisfy their goals. On one hand, this would augment user autonomy by providing an expressive grammar by which to interact with social media. And on the other hand it would spur innovation as developers compete to make newsfeed algorithms and skins that users actually want to adopt. The economics of this paradigm involve unseating the dominance that current SMPs have on the market, which is exciting future work beyond the scope of this white paper. 

Along the lines of this call for interoperable systems are questions of who should own the underlying data. Currently, the data is centrally owned and stored by the SMPs themselves. However, there is a growing body of work that suggests that the data can be owned and stored by the communities that create that data, in data cooperatives \cite{greshake2019open, pentland20202} or a pluralistic social network \citep{pluralistic, zignani2018follow, zulli2020rethinking}.

There are several limitations to the existing instantiation of Yourfeed. First, like its survey antecedents, there is no explicitly ``social'' components of the browsing experience: users have no profile picture or other identifying characteristics, posts have no author, and perhaps most critically, clicking share does not in fact share that post with anyone. These decisions are intentional, as we focused primarily on capturing the attentional affordances of browsing a feed, not the social affordances. However, we hope the next version of Yourfeed accommodates more of these features. In addition, Yourfeed caters to a research audience who recruits participants from online labor markets like Amazon's Mechanical Turk or Lucid. As such, interacting with the platform is part of a transactional task for money, rather than the \textit{in situ} way most people interact with social media. In the future, we hope Yourfeed and tools like it that 1) offer radical alternatives to the social media status quo and 2) prioritize experimentation and public knowledge over profit will become more preferable for social media users, and therefore generate organic participation. 

Building on a legacy of speculative design \citep{dunne2013speculative}, Yourfeed offers a low-friction sandbox for designers, scientists, and citizens to prototype their own ideas for how social media should look and operate. We hope the development of such affirmative visions will bootstrap all three prongs of reform, and in turn help social media become a tool for helping society solve complex challenges.






\bibliographystyle{ACM-Reference-Format}
\bibliography{iccc}


\begin{thebibliography}{47}


\ifx \showCODEN    \undefined \def \showCODEN     #1{\unskip}     \fi
\ifx \showDOI      \undefined \def \showDOI       #1{#1}\fi
\ifx \showISBNx    \undefined \def \showISBNx     #1{\unskip}     \fi
\ifx \showISBNxiii \undefined \def \showISBNxiii  #1{\unskip}     \fi
\ifx \showISSN     \undefined \def \showISSN      #1{\unskip}     \fi
\ifx \showLCCN     \undefined \def \showLCCN      #1{\unskip}     \fi
\ifx \shownote     \undefined \def \shownote      #1{#1}          \fi
\ifx \showarticletitle \undefined \def \showarticletitle #1{#1}   \fi
\ifx \showURL      \undefined \def \showURL       {\relax}        \fi
\providecommand\bibfield[2]{#2}
\providecommand\bibinfo[2]{#2}
\providecommand\natexlab[1]{#1}
\providecommand\showeprint[2][]{arXiv:#2}

\bibitem[\protect\citeauthoryear{Avram, Micallef, Patil, and Menczer}{Avram
  et~al\mbox{.}}{2020}]%
        {avram2020exposure}
\bibfield{author}{\bibinfo{person}{Mihai Avram}, \bibinfo{person}{Nicholas
  Micallef}, \bibinfo{person}{Sameer Patil}, {and} \bibinfo{person}{Filippo
  Menczer}.} \bibinfo{year}{2020}\natexlab{}.
\newblock \showarticletitle{Exposure to social engagement metrics increases
  vulnerability to misinformation}.
\newblock \bibinfo{journal}{\emph{arXiv preprint arXiv:2005.04682}}
  (\bibinfo{year}{2020}).
\newblock


\bibitem[\protect\citeauthoryear{Bail et~al\mbox{.}}{Bail
  et~al\mbox{.}}{2022}]%
        {bail2022social}
\bibfield{author}{\bibinfo{person}{Chris Bail} {et~al\mbox{.}}}
  \bibinfo{year}{2022}\natexlab{}.
\newblock \showarticletitle{Social-media reform is flying blind}.
\newblock \bibinfo{journal}{\emph{Nature}} \bibinfo{volume}{603},
  \bibinfo{number}{7903} (\bibinfo{year}{2022}), \bibinfo{pages}{766--766}.
\newblock


\bibitem[\protect\citeauthoryear{Bhadani, Yamaya, Flammini, Menczer,
  Ciampaglia, and Nyhan}{Bhadani et~al\mbox{.}}{2022}]%
        {bhadani2022political}
\bibfield{author}{\bibinfo{person}{Saumya Bhadani}, \bibinfo{person}{Shun
  Yamaya}, \bibinfo{person}{Alessandro Flammini}, \bibinfo{person}{Filippo
  Menczer}, \bibinfo{person}{Giovanni~Luca Ciampaglia}, {and}
  \bibinfo{person}{Brendan Nyhan}.} \bibinfo{year}{2022}\natexlab{}.
\newblock \showarticletitle{Political audience diversity and news reliability
  in algorithmic ranking}.
\newblock \bibinfo{journal}{\emph{Nature Human Behaviour}} \bibinfo{volume}{6},
  \bibinfo{number}{4} (\bibinfo{year}{2022}), \bibinfo{pages}{495--505}.
\newblock


\bibitem[\protect\citeauthoryear{Bhargava, Chung, Gaikwad, Hope, Jen,
  Rubinovitz, Sald{\'\i}as-Fuentes, and Zuckerman}{Bhargava
  et~al\mbox{.}}{2019}]%
        {bhargava2019gobo}
\bibfield{author}{\bibinfo{person}{Rahul Bhargava}, \bibinfo{person}{Anna
  Chung}, \bibinfo{person}{Neil~S Gaikwad}, \bibinfo{person}{Alexis Hope},
  \bibinfo{person}{Dennis Jen}, \bibinfo{person}{Jasmin Rubinovitz},
  \bibinfo{person}{Bel{\'e}n Sald{\'\i}as-Fuentes}, {and}
  \bibinfo{person}{Ethan Zuckerman}.} \bibinfo{year}{2019}\natexlab{}.
\newblock \showarticletitle{Gobo: A system for exploring user control of
  invisible algorithms in social media}. In
  \bibinfo{booktitle}{\emph{Conference Companion Publication of the 2019 on
  Computer Supported Cooperative Work and Social Computing}}.
  \bibinfo{pages}{151--155}.
\newblock


\bibitem[\protect\citeauthoryear{Bond, Fariss, Jones, Kramer, Marlow, Settle,
  and Fowler}{Bond et~al\mbox{.}}{2012}]%
        {bond201261}
\bibfield{author}{\bibinfo{person}{Robert~M Bond},
  \bibinfo{person}{Christopher~J Fariss}, \bibinfo{person}{Jason~J Jones},
  \bibinfo{person}{Adam~DI Kramer}, \bibinfo{person}{Cameron Marlow},
  \bibinfo{person}{Jaime~E Settle}, {and} \bibinfo{person}{James~H Fowler}.}
  \bibinfo{year}{2012}\natexlab{}.
\newblock \showarticletitle{A 61-million-person experiment in social influence
  and political mobilization}.
\newblock \bibinfo{journal}{\emph{Nature}} \bibinfo{volume}{489},
  \bibinfo{number}{7415} (\bibinfo{year}{2012}), \bibinfo{pages}{295--298}.
\newblock


\bibitem[\protect\citeauthoryear{Butler, Lamont, Law Yim~Wan, Prike, Nasim,
  Walker, Fay, and Ecker}{Butler et~al\mbox{.}}{2022}]%
        {misinfo_game}
\bibfield{author}{\bibinfo{person}{Lucy Butler}, \bibinfo{person}{Padraig
  Lamont}, \bibinfo{person}{Dean Law Yim~Wan}, \bibinfo{person}{Toby Prike},
  \bibinfo{person}{Mehwish Nasim}, \bibinfo{person}{Bradley Walker},
  \bibinfo{person}{Nicolas Fay}, {and} \bibinfo{person}{Ullrich Ecker}.}
  \bibinfo{year}{2022}\natexlab{}.
\newblock \bibinfo{title}{The (Mis)Information Game: A Social Media Simulator}.
\newblock \bibinfo{howpublished}{PsyArxiv}.
\newblock
\urldef\tempurl%
\url{https://psyarxiv.com/628wc/}
\showURL{%
\tempurl}


\bibitem[\protect\citeauthoryear{Cameron, Wodinsky, and DeGeurin}{Cameron
  et~al\mbox{.}}{2022}]%
        {gizmodo}
\bibfield{author}{\bibinfo{person}{Dell Cameron}, \bibinfo{person}{Shoshana
  Wodinsky}, {and} \bibinfo{person}{Mark DeGeurin}.}
  \bibinfo{year}{2022}\natexlab{}.
\newblock \bibinfo{title}{We're Publishing the Facebook Papers. Here's How
  Facebook Killed News Feed Fixes Over Fear of Conservative Backlash.}
\newblock \bibinfo{howpublished}{Web}.
\newblock


\bibitem[\protect\citeauthoryear{Cen and Shah}{Cen and Shah}{2021}]%
        {cen2021regulating}
\bibfield{author}{\bibinfo{person}{Sarah Cen} {and} \bibinfo{person}{Devavrat
  Shah}.} \bibinfo{year}{2021}\natexlab{}.
\newblock \showarticletitle{Regulating algorithmic filtering on social media}.
\newblock \bibinfo{journal}{\emph{Advances in Neural Information Processing
  Systems}}  \bibinfo{volume}{34} (\bibinfo{year}{2021}),
  \bibinfo{pages}{6997--7011}.
\newblock


\bibitem[\protect\citeauthoryear{Chow}{Chow}{2022}]%
        {pluralistic}
\bibfield{author}{\bibinfo{person}{Wes. Chow}.}
  \bibinfo{year}{2022}\natexlab{}.
\newblock \bibinfo{title}{A Sociotechnical Framework for a Pluralistic Social
  Network}.
\newblock \bibinfo{howpublished}{Github}.
\newblock
\urldef\tempurl%
\url{https://github.com/wesc/sf-psn}
\showURL{%
\tempurl}


\bibitem[\protect\citeauthoryear{Doctorow}{Doctorow}{2019}]%
        {ss1}
\bibfield{author}{\bibinfo{person}{Cory Doctorow}.}
  \bibinfo{year}{2019}\natexlab{}.
\newblock \bibinfo{title}{Facebook promised to provide academics data to study
  disinformation, but their foot-dragging has endangered the whole project}.
\newblock \bibinfo{howpublished}{Boingbong}.
\newblock
\urldef\tempurl%
\url{https://boingboing.net/2019/12/12/gdpr-all-purpose-excuse.html}
\showURL{%
\tempurl}


\bibitem[\protect\citeauthoryear{Dunne and Raby}{Dunne and Raby}{2013}]%
        {dunne2013speculative}
\bibfield{author}{\bibinfo{person}{Anthony Dunne} {and} \bibinfo{person}{Fiona
  Raby}.} \bibinfo{year}{2013}\natexlab{}.
\newblock \bibinfo{booktitle}{\emph{Speculative everything: design, fiction,
  and social dreaming}}.
\newblock \bibinfo{publisher}{MIT press}.
\newblock


\bibitem[\protect\citeauthoryear{Eckles}{Eckles}{2022}]%
        {eckles2022algorithmic}
\bibfield{author}{\bibinfo{person}{Dean Eckles}.}
  \bibinfo{year}{2022}\natexlab{}.
\newblock \showarticletitle{Algorithmic transparency and assessing effects of
  algorithmic ranking}.
\newblock  (\bibinfo{year}{2022}).
\newblock


\bibitem[\protect\citeauthoryear{Epstein, Boulais, Gordon, and Groh}{Epstein
  et~al\mbox{.}}{2020}]%
        {epstein2020interpolating}
\bibfield{author}{\bibinfo{person}{Ziv Epstein}, \bibinfo{person}{Oc{\'e}ane
  Boulais}, \bibinfo{person}{Skylar Gordon}, {and} \bibinfo{person}{Matt
  Groh}.} \bibinfo{year}{2020}\natexlab{}.
\newblock \showarticletitle{Interpolating GANs to scaffold autotelic
  creativity}.
\newblock \bibinfo{journal}{\emph{arXiv preprint arXiv:2007.11119}}
  (\bibinfo{year}{2020}).
\newblock


\bibitem[\protect\citeauthoryear{Epstein, Groh, Dubey, and Pentland}{Epstein
  et~al\mbox{.}}{2021}]%
        {epstein2021social}
\bibfield{author}{\bibinfo{person}{Ziv Epstein}, \bibinfo{person}{Matthew
  Groh}, \bibinfo{person}{Abhimanyu Dubey}, {and} \bibinfo{person}{Alex
  Pentland}.} \bibinfo{year}{2021}\natexlab{}.
\newblock \showarticletitle{Social influence leads to the formation of diverse
  local trends}.
\newblock \bibinfo{journal}{\emph{Proceedings of the ACM on Human-Computer
  Interaction}} \bibinfo{volume}{5}, \bibinfo{number}{CSCW2}
  (\bibinfo{year}{2021}), \bibinfo{pages}{1--18}.
\newblock


\bibitem[\protect\citeauthoryear{Epstein, Payne, Shen, Hong, Felbo, Dubey,
  Groh, Obradovich, Cebrian, and Rahwan}{Epstein et~al\mbox{.}}{2018}]%
        {epstein2018turingbox}
\bibfield{author}{\bibinfo{person}{Ziv Epstein}, \bibinfo{person}{Blakeley~H
  Payne}, \bibinfo{person}{Judy~Hanwen Shen}, \bibinfo{person}{Casey~Jisoo
  Hong}, \bibinfo{person}{Bjarke Felbo}, \bibinfo{person}{Abhimanyu Dubey},
  \bibinfo{person}{Matthew Groh}, \bibinfo{person}{Nick Obradovich},
  \bibinfo{person}{Manuel Cebrian}, {and} \bibinfo{person}{Iyad Rahwan}.}
  \bibinfo{year}{2018}\natexlab{}.
\newblock \showarticletitle{TuringBox: An experimental platform for the
  evaluation of AI systems}. In \bibinfo{booktitle}{\emph{IJCAI 2018}}.
  International Joint Conferences on Artificial Intelligence,
  \bibinfo{pages}{5826--5828}.
\newblock


\bibitem[\protect\citeauthoryear{Eslami, Rickman, Vaccaro, Aleyasen, Vuong,
  Karahalios, Hamilton, and Sandvig}{Eslami et~al\mbox{.}}{2015}]%
        {eslami2015always}
\bibfield{author}{\bibinfo{person}{Motahhare Eslami}, \bibinfo{person}{Aimee
  Rickman}, \bibinfo{person}{Kristen Vaccaro}, \bibinfo{person}{Amirhossein
  Aleyasen}, \bibinfo{person}{Andy Vuong}, \bibinfo{person}{Karrie Karahalios},
  \bibinfo{person}{Kevin Hamilton}, {and} \bibinfo{person}{Christian Sandvig}.}
  \bibinfo{year}{2015}\natexlab{}.
\newblock \showarticletitle{" I always assumed that I wasn't really that close
  to [her]" Reasoning about Invisible Algorithms in News Feeds}. In
  \bibinfo{booktitle}{\emph{Proceedings of the 33rd annual ACM conference on
  human factors in computing systems}}. \bibinfo{pages}{153--162}.
\newblock


\bibitem[\protect\citeauthoryear{Gray, Kou, Battles, Hoggatt, and Toombs}{Gray
  et~al\mbox{.}}{2018}]%
        {gray2018dark}
\bibfield{author}{\bibinfo{person}{Colin~M Gray}, \bibinfo{person}{Yubo Kou},
  \bibinfo{person}{Bryan Battles}, \bibinfo{person}{Joseph Hoggatt}, {and}
  \bibinfo{person}{Austin~L Toombs}.} \bibinfo{year}{2018}\natexlab{}.
\newblock \showarticletitle{The dark (patterns) side of UX design}. In
  \bibinfo{booktitle}{\emph{Proceedings of the 2018 CHI conference on human
  factors in computing systems}}. \bibinfo{pages}{1--14}.
\newblock


\bibitem[\protect\citeauthoryear{Greene, Martens, and Shmueli}{Greene
  et~al\mbox{.}}{2022}]%
        {greene2022barriers}
\bibfield{author}{\bibinfo{person}{Travis Greene}, \bibinfo{person}{David
  Martens}, {and} \bibinfo{person}{Galit Shmueli}.}
  \bibinfo{year}{2022}\natexlab{}.
\newblock \showarticletitle{Barriers to academic data science research in the
  new realm of algorithmic behaviour modification by digital platforms}.
\newblock \bibinfo{journal}{\emph{Nature Machine Intelligence}}
  \bibinfo{volume}{4}, \bibinfo{number}{4} (\bibinfo{year}{2022}),
  \bibinfo{pages}{323--330}.
\newblock


\bibitem[\protect\citeauthoryear{Greshake~Tzovaras, Angrist, Arvai, Dulaney,
  Estrada-Gali{\~n}anes, Gunderson, Head, Lewis, Nov, Shaer,
  et~al\mbox{.}}{Greshake~Tzovaras et~al\mbox{.}}{2019}]%
        {greshake2019open}
\bibfield{author}{\bibinfo{person}{Bastian Greshake~Tzovaras},
  \bibinfo{person}{Misha Angrist}, \bibinfo{person}{Kevin Arvai},
  \bibinfo{person}{Mairi Dulaney}, \bibinfo{person}{Vero
  Estrada-Gali{\~n}anes}, \bibinfo{person}{Beau Gunderson},
  \bibinfo{person}{Tim Head}, \bibinfo{person}{Dana Lewis},
  \bibinfo{person}{Oded Nov}, \bibinfo{person}{Orit Shaer}, {et~al\mbox{.}}}
  \bibinfo{year}{2019}\natexlab{}.
\newblock \showarticletitle{Open Humans: A platform for participant-centered
  research and personal data exploration}.
\newblock \bibinfo{journal}{\emph{GigaScience}} \bibinfo{volume}{8},
  \bibinfo{number}{6} (\bibinfo{year}{2019}), \bibinfo{pages}{giz076}.
\newblock


\bibitem[\protect\citeauthoryear{Hao}{Hao}{2021}]%
        {hao2021facebook}
\bibfield{author}{\bibinfo{person}{Karen Hao}.}
  \bibinfo{year}{2021}\natexlab{}.
\newblock \showarticletitle{How Facebook got addicted to spreading
  misinformation}.
\newblock \bibinfo{journal}{\emph{MIT Technology Review}}
  (\bibinfo{year}{2021}).
\newblock


\bibitem[\protect\citeauthoryear{Harris}{Harris}{2016}]%
        {harris}
\bibfield{author}{\bibinfo{person}{Tristan Harris}.}
  \bibinfo{year}{2016}\natexlab{}.
\newblock \bibinfo{title}{How Technology is Hijacking Your Mind — from a
  Magician and Google Design Ethicist}.
\newblock \bibinfo{howpublished}{Medium}.
\newblock


\bibitem[\protect\citeauthoryear{Hegelich}{Hegelich}{2020}]%
        {hegelich2020facebook}
\bibfield{author}{\bibinfo{person}{Simon Hegelich}.}
  \bibinfo{year}{2020}\natexlab{}.
\newblock \showarticletitle{Facebook needs to share more with researchers}.
\newblock \bibinfo{journal}{\emph{Nature}} \bibinfo{volume}{579},
  \bibinfo{number}{7800} (\bibinfo{year}{2020}), \bibinfo{pages}{473--474}.
\newblock


\bibitem[\protect\citeauthoryear{Hwang}{Hwang}{2020}]%
        {hwang2020subprime}
\bibfield{author}{\bibinfo{person}{Tim Hwang}.}
  \bibinfo{year}{2020}\natexlab{}.
\newblock \bibinfo{booktitle}{\emph{Subprime attention crisis: Advertising and
  the time bomb at the heart of the Internet}}.
\newblock \bibinfo{publisher}{FSG originals}.
\newblock


\bibitem[\protect\citeauthoryear{Jagayat, Boparai, Pun, and Choma}{Jagayat
  et~al\mbox{.}}{2021}]%
        {msmwt}
\bibfield{author}{\bibinfo{person}{Arvin Jagayat}, \bibinfo{person}{Gurkaran
  Boparai}, \bibinfo{person}{Carson Pun}, {and} \bibinfo{person}{Becky Choma}.}
  \bibinfo{year}{2021}\natexlab{}.
\newblock \bibinfo{title}{Mock Social Media Website Tool}.
\newblock
\newblock
\urldef\tempurl%
\url{https://docs.studysocial.media}
\showURL{%
\tempurl}


\bibitem[\protect\citeauthoryear{Jouhki, Lauk, Penttinen, Sormanen, and
  Uskali}{Jouhki et~al\mbox{.}}{2016}]%
        {jouhki2016facebook}
\bibfield{author}{\bibinfo{person}{Jukka Jouhki}, \bibinfo{person}{Epp Lauk},
  \bibinfo{person}{Maija Penttinen}, \bibinfo{person}{Niina Sormanen}, {and}
  \bibinfo{person}{Turo Uskali}.} \bibinfo{year}{2016}\natexlab{}.
\newblock \showarticletitle{Facebook’s emotional contagion experiment as a
  challenge to research ethics}.
\newblock \bibinfo{journal}{\emph{Media and Communication}}
  \bibinfo{volume}{4} (\bibinfo{year}{2016}).
\newblock


\bibitem[\protect\citeauthoryear{Kramer, Guillory, and Hancock}{Kramer
  et~al\mbox{.}}{2014}]%
        {kramer2014experimental}
\bibfield{author}{\bibinfo{person}{Adam~DI Kramer}, \bibinfo{person}{Jamie~E
  Guillory}, {and} \bibinfo{person}{Jeffrey~T Hancock}.}
  \bibinfo{year}{2014}\natexlab{}.
\newblock \showarticletitle{Experimental evidence of massive-scale emotional
  contagion through social networks}.
\newblock \bibinfo{journal}{\emph{Proceedings of the National Academy of
  Sciences}} \bibinfo{volume}{111}, \bibinfo{number}{24}
  (\bibinfo{year}{2014}), \bibinfo{pages}{8788--8790}.
\newblock


\bibitem[\protect\citeauthoryear{Lazer, Pentland, Watts, Aral, Athey,
  Contractor, Freelon, Gonzalez-Bailon, King, Margetts, et~al\mbox{.}}{Lazer
  et~al\mbox{.}}{2020}]%
        {lazer2020computational}
\bibfield{author}{\bibinfo{person}{David~MJ Lazer}, \bibinfo{person}{Alex
  Pentland}, \bibinfo{person}{Duncan~J Watts}, \bibinfo{person}{Sinan Aral},
  \bibinfo{person}{Susan Athey}, \bibinfo{person}{Noshir Contractor},
  \bibinfo{person}{Deen Freelon}, \bibinfo{person}{Sandra Gonzalez-Bailon},
  \bibinfo{person}{Gary King}, \bibinfo{person}{Helen Margetts},
  {et~al\mbox{.}}} \bibinfo{year}{2020}\natexlab{}.
\newblock \showarticletitle{Computational social science: Obstacles and
  opportunities}.
\newblock \bibinfo{journal}{\emph{Science}} \bibinfo{volume}{369},
  \bibinfo{number}{6507} (\bibinfo{year}{2020}), \bibinfo{pages}{1060--1062}.
\newblock


\bibitem[\protect\citeauthoryear{Lin, Werner, and Inzlicht}{Lin
  et~al\mbox{.}}{2021}]%
        {lin2021promises}
\bibfield{author}{\bibinfo{person}{Hause Lin}, \bibinfo{person}{Kaitlyn~M
  Werner}, {and} \bibinfo{person}{Michael Inzlicht}.}
  \bibinfo{year}{2021}\natexlab{}.
\newblock \showarticletitle{Promises and perils of experimentation: The
  mutual-internal-validity problem}.
\newblock \bibinfo{journal}{\emph{Perspectives on Psychological Science}}
  \bibinfo{volume}{16}, \bibinfo{number}{4} (\bibinfo{year}{2021}),
  \bibinfo{pages}{854--863}.
\newblock


\bibitem[\protect\citeauthoryear{Lorenz-Spreen, Lewandowsky, Sunstein, and
  Hertwig}{Lorenz-Spreen et~al\mbox{.}}{2020}]%
        {lorenz2020behavioural}
\bibfield{author}{\bibinfo{person}{Philipp Lorenz-Spreen},
  \bibinfo{person}{Stephan Lewandowsky}, \bibinfo{person}{Cass~R Sunstein},
  {and} \bibinfo{person}{Ralph Hertwig}.} \bibinfo{year}{2020}\natexlab{}.
\newblock \showarticletitle{How behavioural sciences can promote truth,
  autonomy and democratic discourse online}.
\newblock \bibinfo{journal}{\emph{Nature human behaviour}} \bibinfo{volume}{4},
  \bibinfo{number}{11} (\bibinfo{year}{2020}), \bibinfo{pages}{1102--1109}.
\newblock


\bibitem[\protect\citeauthoryear{Mahajan, Roy~Choudhury, Levens, Gallicano, and
  Shaikh}{Mahajan et~al\mbox{.}}{2021}]%
        {mahajan2021community}
\bibfield{author}{\bibinfo{person}{Khyati Mahajan}, \bibinfo{person}{Sourav
  Roy~Choudhury}, \bibinfo{person}{Sara Levens}, \bibinfo{person}{Tiffany
  Gallicano}, {and} \bibinfo{person}{Samira Shaikh}.}
  \bibinfo{year}{2021}\natexlab{}.
\newblock \showarticletitle{Community connect: A mock social media platform to
  study online behavior}. In \bibinfo{booktitle}{\emph{Proceedings of the 14th
  ACM International Conference on Web Search and Data Mining}}.
  \bibinfo{pages}{1073--1076}.
\newblock


\bibitem[\protect\citeauthoryear{Mosleh, Pennycook, and Rand}{Mosleh
  et~al\mbox{.}}{2020}]%
        {mosleh2020self}
\bibfield{author}{\bibinfo{person}{Mohsen Mosleh}, \bibinfo{person}{Gordon
  Pennycook}, {and} \bibinfo{person}{David~G Rand}.}
  \bibinfo{year}{2020}\natexlab{}.
\newblock \showarticletitle{Self-reported willingness to share political news
  articles in online surveys correlates with actual sharing on Twitter}.
\newblock \bibinfo{journal}{\emph{Plos one}} \bibinfo{volume}{15},
  \bibinfo{number}{2} (\bibinfo{year}{2020}), \bibinfo{pages}{e0228882}.
\newblock


\bibitem[\protect\citeauthoryear{Mosleh, Pennycook, and Rand}{Mosleh
  et~al\mbox{.}}{2022}]%
        {mosleh2022field}
\bibfield{author}{\bibinfo{person}{Mohsen Mosleh}, \bibinfo{person}{Gordon
  Pennycook}, {and} \bibinfo{person}{David~G Rand}.}
  \bibinfo{year}{2022}\natexlab{}.
\newblock \showarticletitle{Field experiments on social media}.
\newblock \bibinfo{journal}{\emph{Current Directions in Psychological Science}}
  \bibinfo{volume}{31}, \bibinfo{number}{1} (\bibinfo{year}{2022}),
  \bibinfo{pages}{69--75}.
\newblock


\bibitem[\protect\citeauthoryear{Oremus, Alcantara, Merrill, and
  Galocha}{Oremus et~al\mbox{.}}{2021}]%
        {fb_wpo}
\bibfield{author}{\bibinfo{person}{Will Oremus}, \bibinfo{person}{Chris
  Alcantara}, \bibinfo{person}{Jeremy Merrill}, {and} \bibinfo{person}{Artur
  Galocha}.} \bibinfo{year}{2021}\natexlab{}.
\newblock \bibinfo{title}{How Facebook shapes your feed: The evolution of what
  posts get top billing on users’ news feeds, and what gets obscured}.
\newblock \bibinfo{howpublished}{Web}.
\newblock
\urldef\tempurl%
\url{https://www.washingtonpost.com/technology/interactive/2021/how-facebook-algorithm-works/}
\showURL{%
\tempurl}


\bibitem[\protect\citeauthoryear{Pennycook, Epstein, Mosleh, Arechar, Eckles,
  and Rand}{Pennycook et~al\mbox{.}}{2021}]%
        {pennycook2021shifting}
\bibfield{author}{\bibinfo{person}{Gordon Pennycook}, \bibinfo{person}{Ziv
  Epstein}, \bibinfo{person}{Mohsen Mosleh}, \bibinfo{person}{Antonio~A
  Arechar}, \bibinfo{person}{Dean Eckles}, {and} \bibinfo{person}{David~G
  Rand}.} \bibinfo{year}{2021}\natexlab{}.
\newblock \showarticletitle{Shifting attention to accuracy can reduce
  misinformation online}.
\newblock \bibinfo{journal}{\emph{Nature}} \bibinfo{volume}{592},
  \bibinfo{number}{7855} (\bibinfo{year}{2021}), \bibinfo{pages}{590--595}.
\newblock


\bibitem[\protect\citeauthoryear{Pennycook and Rand}{Pennycook and
  Rand}{2021}]%
        {pennycook2021psychology}
\bibfield{author}{\bibinfo{person}{Gordon Pennycook} {and}
  \bibinfo{person}{David~G Rand}.} \bibinfo{year}{2021}\natexlab{}.
\newblock \showarticletitle{The psychology of fake news}.
\newblock \bibinfo{journal}{\emph{Trends in cognitive sciences}}
  \bibinfo{volume}{25}, \bibinfo{number}{5} (\bibinfo{year}{2021}),
  \bibinfo{pages}{388--402}.
\newblock


\bibitem[\protect\citeauthoryear{Pentland and Hardjono}{Pentland and
  Hardjono}{2020}]%
        {pentland20202}
\bibfield{author}{\bibinfo{person}{Alex Pentland} {and} \bibinfo{person}{Thomas
  Hardjono}.} \bibinfo{year}{2020}\natexlab{}.
\newblock \showarticletitle{2. Data Cooperatives}.
\newblock In \bibinfo{booktitle}{\emph{Building the New Economy}}.
  \bibinfo{publisher}{PubPub}.
\newblock


\bibitem[\protect\citeauthoryear{Reuning, Whitesell, and Hannah}{Reuning
  et~al\mbox{.}}{2022}]%
        {reuning2022facebook}
\bibfield{author}{\bibinfo{person}{Kevin Reuning}, \bibinfo{person}{Anne
  Whitesell}, {and} \bibinfo{person}{A~Lee Hannah}.}
  \bibinfo{year}{2022}\natexlab{}.
\newblock \showarticletitle{Facebook algorithm changes may have amplified local
  republican parties}.
\newblock \bibinfo{journal}{\emph{Research \& Politics}} \bibinfo{volume}{9},
  \bibinfo{number}{2} (\bibinfo{year}{2022}),
  \bibinfo{pages}{20531680221103809}.
\newblock


\bibitem[\protect\citeauthoryear{Roozenbeek and Van~der Linden}{Roozenbeek and
  Van~der Linden}{2019}]%
        {roozenbeek2019fake}
\bibfield{author}{\bibinfo{person}{Jon Roozenbeek} {and}
  \bibinfo{person}{Sander Van~der Linden}.} \bibinfo{year}{2019}\natexlab{}.
\newblock \showarticletitle{Fake news game confers psychological resistance
  against online misinformation}.
\newblock \bibinfo{journal}{\emph{Palgrave Communications}}
  \bibinfo{volume}{5}, \bibinfo{number}{1} (\bibinfo{year}{2019}),
  \bibinfo{pages}{1--10}.
\newblock


\bibitem[\protect\citeauthoryear{Sadowski, Viljoen, and Whittaker}{Sadowski
  et~al\mbox{.}}{2021}]%
        {sadowski2021everyone}
\bibfield{author}{\bibinfo{person}{Jathan Sadowski},
  \bibinfo{person}{Salom{\'e} Viljoen}, {and} \bibinfo{person}{Meredith
  Whittaker}.} \bibinfo{year}{2021}\natexlab{}.
\newblock \bibinfo{title}{Everyone should decide how their digital data are
  used—Not just tech companies}.
\newblock
\newblock


\bibitem[\protect\citeauthoryear{Salganik, Dodds, and Watts}{Salganik
  et~al\mbox{.}}{2006}]%
        {salganik2006experimental}
\bibfield{author}{\bibinfo{person}{Matthew~J Salganik},
  \bibinfo{person}{Peter~Sheridan Dodds}, {and} \bibinfo{person}{Duncan~J
  Watts}.} \bibinfo{year}{2006}\natexlab{}.
\newblock \showarticletitle{Experimental study of inequality and
  unpredictability in an artificial cultural market}.
\newblock \bibinfo{journal}{\emph{science}} \bibinfo{volume}{311},
  \bibinfo{number}{5762} (\bibinfo{year}{2006}), \bibinfo{pages}{854--856}.
\newblock


\bibitem[\protect\citeauthoryear{Sassine, Rahimian, and Eckles}{Sassine
  et~al\mbox{.}}{2022}]%
        {sassine2022influence}
\bibfield{author}{\bibinfo{person}{Jad Sassine}, \bibinfo{person}{M~Amin
  Rahimian}, {and} \bibinfo{person}{Dean Eckles}.}
  \bibinfo{year}{2022}\natexlab{}.
\newblock \showarticletitle{Influence of Repetition through Limited Recall}. In
  \bibinfo{booktitle}{\emph{Proceedings of the International AAAI Conference on
  Web and Social Media}}, Vol.~\bibinfo{volume}{16}. \bibinfo{pages}{863--872}.
\newblock


\bibitem[\protect\citeauthoryear{Smith}{Smith}{2021}]%
        {tiktok_nyt}
\bibfield{author}{\bibinfo{person}{Ben Smith}.}
  \bibinfo{year}{2021}\natexlab{}.
\newblock \bibinfo{title}{How TikTok Reads Your Mind}.
\newblock \bibinfo{howpublished}{Web}.
\newblock
\urldef\tempurl%
\url{https://www.nytimes.com/2021/12/05/business/media/tiktok-algorithm.html}
\showURL{%
\tempurl}


\bibitem[\protect\citeauthoryear{Team}{Team}{2021}]%
        {tiktok_wsj}
\bibfield{author}{\bibinfo{person}{WSJ Team}.} \bibinfo{year}{2021}\natexlab{}.
\newblock \bibinfo{title}{Inside TikTok’s Algorithm: A WSJ Video
  Investigation}.
\newblock \bibinfo{howpublished}{Web}.
\newblock
\urldef\tempurl%
\url{https://www.wsj.com/articles/tiktok-algorithm-video-investigation-11626877477}
\showURL{%
\tempurl}


\bibitem[\protect\citeauthoryear{Zignani, Gaito, and Rossi}{Zignani
  et~al\mbox{.}}{2018}]%
        {zignani2018follow}
\bibfield{author}{\bibinfo{person}{Matteo Zignani}, \bibinfo{person}{Sabrina
  Gaito}, {and} \bibinfo{person}{Gian~Paolo Rossi}.}
  \bibinfo{year}{2018}\natexlab{}.
\newblock \showarticletitle{Follow the “mastodon”: Structure and evolution
  of a decentralized online social network}. In
  \bibinfo{booktitle}{\emph{Proceedings of the International AAAI Conference on
  Web and Social Media}}, Vol.~\bibinfo{volume}{12}. \bibinfo{pages}{541--550}.
\newblock


\bibitem[\protect\citeauthoryear{Zittrain}{Zittrain}{2014}]%
        {zittrain2014facebook}
\bibfield{author}{\bibinfo{person}{Jonathan Zittrain}.}
  \bibinfo{year}{2014}\natexlab{}.
\newblock \showarticletitle{Facebook could decide an election without anyone
  ever finding out}.
\newblock \bibinfo{journal}{\emph{New Republic}}  \bibinfo{volume}{1}
  (\bibinfo{year}{2014}).
\newblock


\bibitem[\protect\citeauthoryear{Zuboff}{Zuboff}{2015}]%
        {zuboff2015big}
\bibfield{author}{\bibinfo{person}{Shoshana Zuboff}.}
  \bibinfo{year}{2015}\natexlab{}.
\newblock \showarticletitle{Big other: surveillance capitalism and the
  prospects of an information civilization}.
\newblock \bibinfo{journal}{\emph{Journal of information technology}}
  \bibinfo{volume}{30}, \bibinfo{number}{1} (\bibinfo{year}{2015}),
  \bibinfo{pages}{75--89}.
\newblock


\bibitem[\protect\citeauthoryear{Zulli, Liu, and Gehl}{Zulli
  et~al\mbox{.}}{2020}]%
        {zulli2020rethinking}
\bibfield{author}{\bibinfo{person}{Diana Zulli}, \bibinfo{person}{Miao Liu},
  {and} \bibinfo{person}{Robert Gehl}.} \bibinfo{year}{2020}\natexlab{}.
\newblock \showarticletitle{Rethinking the “social” in “social media”:
  Insights into topology, abstraction, and scale on the Mastodon social
  network}.
\newblock \bibinfo{journal}{\emph{New Media \& Society}} \bibinfo{volume}{22},
  \bibinfo{number}{7} (\bibinfo{year}{2020}), \bibinfo{pages}{1188--1205}.
\newblock


\end{thebibliography}

\end{document}